\RequirePackage[2020-02-02]{latexrelease}
\documentclass[twocolumn,english,aps,prl,10pt,superscriptaddress]{revtex4}
\usepackage[T1]{fontenc}
\usepackage{amsmath,graphicx,amssymb,epsfig,dsfont}
\usepackage{mathrsfs}
\usepackage{color}
\usepackage{amsbsy}

\renewcommand{\Im}{{\rm Im}}
\newcommand{\Tr}{{\rm Tr}}
\newcommand{\rd}{{\rm d}}
\newcommand{\kb}{k_{\rm B}}

\def\bbm[#1]{\mbox{\boldmath $#1$}}
\begin{document}

\title{Inverse Spin Thermal Hall Effect in Non-Reciprocal Photonic Systems}

\author{P. Ben-Abdallah}
\email{pba@institutoptique.fr}
\affiliation{Laboratoire Charles Fabry, UMR 8501, Institut d'Optique, CNRS, Universit\'{e} Paris-Saclay, 2 Avenue Augustin Fresnel, 91127 Palaiseau Cedex, France.}

\date{\today}

\begin{abstract}
A transverse radiative heat flux induced by the gradient of spin angular momentum of photons in non-reciprocal systems is predicted. This thermal analog of the inverse spin Hall effect is analyzed in magneto-optical networks exhibiting C4 symmetry, under the action of spatially variable external magnetic fields. This finding opens new avenues for thermal management and energy conversion with non-reciprocal systems through a localized and dynamic control of the spin angular momentum of light.
\end{abstract}

\maketitle

The spin Hall effect (SHE)~\cite{Dyakonov,Hirsch,Kato} and its reciprocal counterpart, the inverse spin Hall effect (ISHE)~\cite{Saitoh,Valenzuela,Zhao,Werake,Sinova} have undisputably paved the way to the modern spintronics by taking advantage of coupling between charge currents and spin dynamics in materials. Recently, after the discovering of singular radiative transport phenomena in non-reciprocal materials such as the photon thermal Hall effect~\cite{pba2016} and the presence of persistent heat currents~\cite{Fan} in systems at equilibrium, a thermal spin photonics (TSP) has emerged ~\cite{Ott1, Jacob1,Fan2,Jacob2} based on the link between the heat flux transported by thermal photons inside non-reciprocal systems, and their angular momentum. This TSP offers possibilities for an unprecedent control of angular momentum of photons emitted by thermal sources.
In this Letter, we predict the existence of an inverse thermal Hall effect driven by the gradient of spin angular momentum in non-reciprocal systems paving the way to a thermal management assisted by a local control of spin angular momentum of light. To highight this effect, also called inverse spin thermal Hall effect (ISTHE), we generate a gradient of spin angular momentum in a regular network of magneto-optical nanoparticles interacting by radiation under the action of an inhomogeneous external magnetic field. Then, we calculate the equilibrium temperature of particles in the direction perpendicular to this primary  gradient to highlight the presence of a transerval temperature gradient.

\begin{figure}
\centering
\includegraphics[angle=0,scale=0.34]{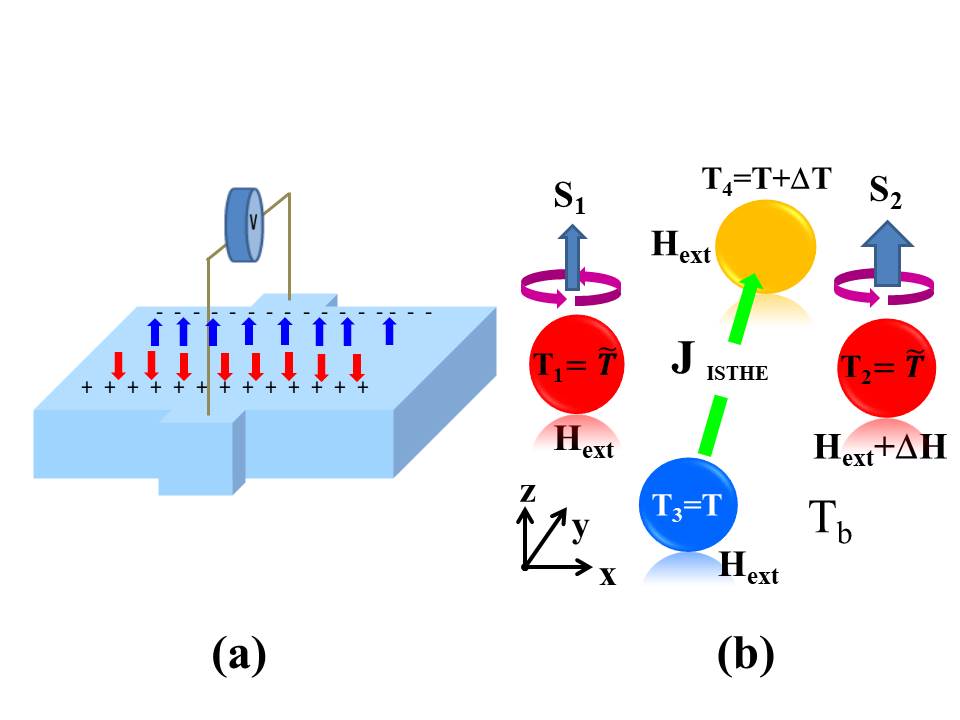}
\caption{Schematic of the:  (a) conventional inverse spin Hall effect (ISHE). A spin current through a material generates a transverse charge current; (b) inverse spin thermal Hall effect (ISTHE) in a non-reciprocal network of C4 symmetry made with  magneto-optical nanoparticles  under spatially variable external magnetic field $\bold{H}_{ext}$ along the $\bold{x}$-direction where no temperature gradient exists (i.e. $T_1=T_2=\tilde{T}$). This spatial variation of external field induces a variation in the same direction of light's spin angular momentum $\mathbf{S}_i$ which, in turn, gives rise to a temperture gradient in the transverse direction (i.e. $T_4=T_3+\Delta T$) and a lateral flux $J_{ISTHE}$. The network is immersed in a thermal bath at temperature $T_b$ and the particles along which a variation of magnetic field is applied are thermostated at a temperature $T\neq T_b$ .} 
\end{figure}

To start, we consider, as for the demonstratrion of photon thermal Hall effect~\cite{pba2016}, a system (Fig.~1) of four identical spherical nanoparticles of radius $R$ made with Indium Antmonide (InSb) a magneto-optical material  which form a network of perfect symmetry C4 immersed in a thermal bath at temperature $T_b$.  We assume the two particles along the $\mathbf{x}$-axis connected to a thermal reservoir at temperature $T_1=T_2=\tilde{T}=T_b+\Delta T$ while the two other particles are left free to equilibrate at  their own equilibrium temperature $T_3$ and $T_4$ with respect to the environment temperature and to the external magnetic field $\mathbf{H}_{ext}(\mathbf{r})$ applied on the system. 
When a uniform  field is applied on the whole system orthogonally to the particles network, no Hall flux can exist~\cite{pba2016} and $T_3=T_4$. On the other hand, as we will see hereafter, when an inhomogeneous magnetic field is applied through the system, the two unthermostated particles reach two different equilibrium temperatures. Therefore a heat flux is exchanged transversally to the gradient of magnetic field.
Using the Landauer formalism for N-body systems the net thermal power received by the $i^{th}$ particle through the radiative interaction reads~\cite{pba2011,biehs1}
\begin{equation}
\varphi_{i}=\underset{j\neq i}{\sum}\varphi_{ji}(\mathbf{H}_{ext})+\varphi_{bi}(\mathbf{H}_{ext})\label{Eq:InterpartHeatFlux},
\end{equation}
where
\begin{equation}
\varphi_{ji} = \int_{0}^{\infty} \frac{\mathrm{d} \omega}{2\pi} \left[ \Theta_{\omega}(T_{j}) \mathcal{T}_{j,i}(\omega, \mathbf{H}_{ext}) - \Theta_{\omega}(T_{i}) \mathcal{T}_{i,j}(\omega, \mathbf{H}_{ext}) \right]
\end{equation}
is the net power exchanged between  the two particles $j$ and $i$ under the action of external magnetic field $\mathbf{H}_{ext}(\mathbf{r})$ while
\begin{equation}
\varphi_{bi}(\mathbf{H}_{ext})=\int_{0}^{\infty}\frac{\rd\omega}{2\pi}\,[\Theta_{\omega}(T_{b})-\Theta_{\omega}(T_{i})]\mathcal{T}_{b,i}(\omega,\mathbf{H}_{ext})
\end{equation}
is the power it exchanges with the external bath.
In these expressions $\Theta_{\omega}(T)={\hbar\omega}/[{e^{\frac{\hbar\omega}{k_B T}}-1}]$ is the mean energy of a harmonic oscillator in
thermal equilibrium at temperature $T$ and $\mathcal{T}_{i,j}(\omega,\mathbf{H}_{ext})$ denotes the transmission coefficient, at the frequency $\omega$, under the action of external magnetic field $\mathbf{H}_{ext}(\mathbf{r})$ between the two particles $i$ and $j$ while $\mathcal{T}_{b,i}=\underset{j}{\sum}\mathcal{T}_{j,i}$ denotes the transmission coefficient between the external bath and the particle $i$.
When the particles are small enough 
compared with their thermal wavelength $\lambda_{T_{i}} = c\hbar/(\kb T_{i})$ ($c$ is the vacuum light
velocity, $2 \pi \hbar$ is Planck's constant, and $\kb$ is Boltzmann's constant) they can be modeled by simple radiating dipoles.
In this case the transmission coefficient is defined as~\cite{Cuevas,Cuevas2}
\begin{equation}
\mathcal{T}_{i,j}(\omega)=\frac{4}{3}(\frac{\omega}{c})^4\Im\Tr\bigl[\boldsymbol{\alpha}_j\mathds{G}^{EE}_{ji}\frac{1}{2i}(\boldsymbol{\alpha}_i-\boldsymbol{\alpha}_i^{\dagger})\mathds{G}^{EE \dagger}_{ji}\bigr],
\end{equation}
where $\boldsymbol{\alpha}_i$ denotes the polarizability tensor of the $i^{th}$ particle and $\mathds{G}^{EE}_{ij}$ is the full dyadic electric-electric Green tensor between the $i^{th}$ and the $j^{th}$ particle which reads
\begin{equation}
  \mathds{G}^{EE}(\bold{r},\bold{r_j})=\underset{i=1}{\overset{N}{\sum}}\mathscr{G}^{EE}_0(\bold{r},\bold{r_i})\mathds{T}_{EE,ij}^{-1}.
  \label{full_Green_electric}
\end{equation}
in terms of free space propagator
\begin{equation}
\begin{split}
\mathscr{G}^{EE}_0(\bold{r'},\bold{r''})=\:\:\:\:\:\:\:\:\:\:\:\:\:\:\:\:\:\:\:\:\:\:\:\:\:\:\:\:\:\:\:\:\:\:\:\:\:\:\:\:\:\:\:\:\:\:\:\:\:\:\:\:\:\:\:\:\:\\
\frac{\exp(ik_0 r)}{4\pi r}\left[\left(1+\frac{ik_0 r-1}{k_0^{2}r^{2}}\right)\mathds{1}+\frac{3-3ik_0 r-k_0^{2}r^{2}}{k_0^{2}r^{2}}\widehat{\mathbf{r}}\otimes\widehat{\mathbf{r}}\right],
\label{propagator_vacuum}
\end{split}
\end{equation}
with $\widehat{\mathbf{r}}\equiv\mathbf{\bold{r}}/r$,
$\mathbf{r}=\bold{r'}-\bold{r''}$ and  $r=\mid\mathbf{r}\mid$ and $\mathds{1}$
 the unit dyadic tensor and in term of the $\mathds{T}_{EE}$  block matrix of component
\begin{equation}
\mathds{T}_{EE,ij} =\delta_{ij}\mathds{1}-(1-\delta_{ij})k_0^2 \boldsymbol{\alpha}_{i} \mathscr{G}^{EE}_0(\bold{r_i},\bold{r_j}).
\label{Eq:A0}
\end{equation}
As, the polarizability tensor is concerned,  it can be described, by taking into account the radiative corrections, using the following expression~\cite{Albaladejo}
\begin{equation}
\boldsymbol{\alpha}_{i}(\omega)=\left( \mathds{1}-i\frac{k^3}{6\pi} \boldsymbol{\alpha}_{0,i}(\omega,)\right)^{-1} \boldsymbol{\alpha}_{0,i}(\omega)\label{Eq:Polarizability2},
\end{equation}
where $ \boldsymbol{\alpha}_{0,i}$ denotes the quasistatic polarizability of the $i^{th}$ particle and $k=\omega/c$, $c$ being the speed of light in vacuum. For spherical particles in vacuum, the quasistatic polarizability takes the simple form
\begin{equation}
 \boldsymbol{\alpha}_{0,i}(\omega)=4\pi R^3\big(\boldsymbol{\varepsilon}_i(\omega)-\mathds{1}\big)\big(\boldsymbol{\varepsilon}_i(\omega)+2\mathds{1}\big)^{-1}\label{Eq:Polarizability2},
\end{equation}
where $\boldsymbol{\varepsilon}_i(\omega)\equiv\boldsymbol{\varepsilon}(\omega,\bold{H}_{ext}(\bold{r}_i))$ is the permittivity of the particle. When a magnetic field is applied in the direction parallel to the $\mathbf{z}$-axis, the permittivity tensor of  particles takes the following form~\cite{Palik,Moncada}
\begin{equation}
\boldsymbol{\varepsilon}=\left(\begin{array}{ccc}
\varepsilon_{1} & -i\varepsilon_{2} & 0\\
i\varepsilon_{2} & \varepsilon_{1} & 0\\
0 & 0 & \varepsilon_{3}
\end{array}\right)\label{Eq:permittivity}
\end{equation}
with
\begin{equation}
\varepsilon_{1}(H_{ext})=\varepsilon_\infty(1+\frac{\omega_L^2-\omega_T^2}{\omega_T^2-\omega^2-i\Gamma\omega}+\frac{\omega_p^2(\omega+i\gamma)}{\omega[\omega_c^2-(\omega+i\gamma)^2]}) \label{Eq:permittivity1},
\end{equation}
\begin{equation}
\varepsilon_{2}(H_{ext})=\frac{\varepsilon_\infty\omega_p^2\omega_c}{\omega[(\omega+i\gamma)^2-\omega_c^2]}\label{Eq:permittivity2},
\end{equation}
\begin{equation}
\varepsilon_{3}=\varepsilon_\infty(1+\frac{\omega_L^2-\omega_T^2}{\omega_T^2-\omega^2-i\Gamma\omega}-\frac{\omega_p^2}{\omega(\omega+i\gamma)})\label{Eq:permittivity3}.
\end{equation}
Here, $\varepsilon_\infty=15.7$ is the infinite-frequency dielectric constant, $\omega_L=3.62\times10^{13} rad.s^{-1}$ is the longitudinal opical phonon frequency, $\omega_T=3.39\times10^{13} rad.s^{-1}$ is the transverse optical phonon frequency, $\omega_p=(\frac{ne^2}{m^*\varepsilon_0\varepsilon_\infty})^{1/2}$ is the plasma frequency of free carriers of density $n=1.36\times10^{19} cm^{-3}$ and effective mass $m^*=7.29\times 10^{-32}kg$, $\Gamma=5.65\times10^{11} rad.s^{-1}$ is the phonon damping constant,$\gamma=1\times10^{12} rad.s^{-1}$ is the free carrier damping constant and $\omega_c=eH_{ext}/m^*$ is the cyclotron frequency. 
Now we assume the external magnetic undergoes a variation $\Delta H$ between the particle $1$ where $\bold{H}_{ext}(\bold{r}_1)=H_{ext}\bold{z}$ and particle $2$ where  $\bold{H}_{ext}(\bold{r}_2)=(H_{ext}+\Delta H)\bold{z}$ while $\bold{H}_{ext}(\bold{r}_3)=\bold{H}_{ext}(\bold{r}_4)=\bold{H}_{ext}(\bold{r}_1)$.
The temperatures of particles $3$ and $4$ can be found from the global equilibrium conditions
\begin{equation}
\varphi_{i}(\bold{H}_{ext};T_3,T_4)=0, \:\:\: i=3,4\label{Eq:equilibrium}.
\end{equation}
In the linear approximation this system can be recasted under the form 
\begin{equation}
\begin{split}
\underset{j\neq i}{\sum}G^{TT}_{ji}(T_j-T_i)+G^{TT}_{bi}(T_b-T_i)\\
+G^{TH}_{2i}\Delta H=0,\:\:\: i=3,4\label{Eq:linear}
\end{split}
\end{equation}
where $G^{TT}_{ji}=\frac{\partial\varphi_{ji}}{\partial T}$ and $G^{TT}_{bi}=\frac{\partial\varphi_{bi}}{\partial T}$ denote the thermal conductances induces by the temperature gradient between the particles $j$ and $i$ and between the bath and the particle $i$, respectively while $G^{TH}_{ji}=\frac{\partial\varphi_{ji}}{\partial H}$ is the thermal conductance associated to the gradient of magnetic field. 
After solving the  linear system (\ref{Eq:linear}) it is straighforward to show that the temperature difference $\Delta T\equiv T_3-T_4$ reads withe respect to the lateral gradient $\Delta H$ of magnetic field and with respect to  the temperature gradient $\delta T=\tilde{T}-T_b$ between the pair of particles 1 and 2 and the thermal bath
\begin{equation}
\Delta T=a_{TT}\delta T+b_{TH}\Delta H \label{temp_diff}
\end{equation}
where
\begin{equation}
a_{TT}=\frac{1}{\Gamma}[G^{TT}_{b4}(G^{TT}_{31}+G^{TT}_{32})-G^{TT}_{b3}(G^{TT}_{41}+G^{TT}_{42})], \label{coeff1}
\end{equation}
\begin{equation}
\begin{split}
b_{TH}=\frac{1}{\Gamma}[G^{TH}_{23}(G^{TT}_{41}+G^{TT}_{42}+G^{TT}_{b4})\\
-G^{TH}_{24}(G^{TT}_{31}+G^{TT}_{32}+G^{TT}_{b3})], \label{coeff2}
\end{split}
\end{equation}
with $\Gamma=(\underset{j\neq 3}{\sum}G^{TT}_{3j}+G^{TT}_{b3})(\underset{j\neq 4}{\sum}G^{TT}_{4j}+G^{TT}_{b4})-G^{TT}_{34}G^{TT}_{43}$. Expression (\ref{temp_diff}) shows that $\Delta T=0$ when $\Delta H=0$ (uniform magnetic field) and $\delta T=0$ (system at thermal equilibrium). On the other hand when $\Delta H\neq 0$ a temperature difference generally exists between the particles $3$ and $4$ even when $\tilde{T}=T_b$ as shown in Fig.2(b). Notice also that, a temperature gradient between the bath and both particles $1$ and $2$ is sufficient to assure the presence of a lateral temperature gradient.
This  temperature difference $\Delta T$ is plotted in Figs. 2  in a regular square of InSb nanoparticles of radius $R=50\:nm$ when $\tilde{T}=310\:K$ and $T_b=300\:K$ (Fig.2(a)) for different value of $\Delta H$ using the solutions of nonlinear system (\ref{Eq:equilibrium}) and when $T_1=T_2=T_b$ (Fig.2(b)). We see that $\Delta T$ is positive for small value of $H_{ext}$ and becomes negative when the magnitude of external magnetic field increases. We also note that the temperature difference is very sensitive to the gradient of this field through the network.
 It is important to stress that the presence of this latteral temperature gradient is fundamentally different from the photon thermal Hall effect~\cite{pba2016} which requires the existence of a temperature gradient in the $\bold{x}$-direction. The plots  in the insets of  Figs.2 show the temperature of particle $3$ and it demonstrates that both particles $3$ and $4$ equilibrate at different temperatures  between $\tilde{T}$ and $T_b$. This temperature difference  is responsible for a permanent exchange of energy $\varphi_{ISTHE}\equiv\varphi_{34}$  between these particles as shown in Figs.3 , the net power received by each of these particles from all surronding particles and from the external bath being obviously always zero in steady state regime. Also, we see that this transversal echange of energy displays a non trivial behavior with respect to the magnetic field $H_{ext}$ applied on the system,  changing even sign several times with respect to the magnitude of this field. When $\varphi_{ISTHE}$ becomes negative and $\Delta T>0$ (resp. $\varphi_{ISTHE}>0$) and $\Delta T<0$)  this demonstrates  that the external magnetic field induces a pumping effect.  This pumping exists even when both particles $1$ and $2$ are set at the same temperature as the thermal bath which shows the fundamentally different nature of this effect compared to the photon thermal Hall effect. Note that the shift of the resonance peaks in Fig. 3(a) to higher frequencies compared to those in Fig. 3(b) results from two things. First, the magneto-dependent resonance frequencies of the particles exhibit a linear dependence on the amplitude of the applied external field (see ref~\cite{pba2016}). Second, according to the Wien's displacement law, the Wien's frequency increases with the temperature of the particles, leading to more effective excitation of resonances when this frequency aligns with the particle's resonance frequency. As a result, when a non-zero temperature gradient $\delta T$ is applied between the pair of particles $1$ and $2$ and the external bath, particles $3$ and $4$ heat up relative to the thermal bath, thereby activating resonances associated with higher field strengths.

\begin{figure}
\centering
\includegraphics[angle=0,scale=0.32,angle=0]{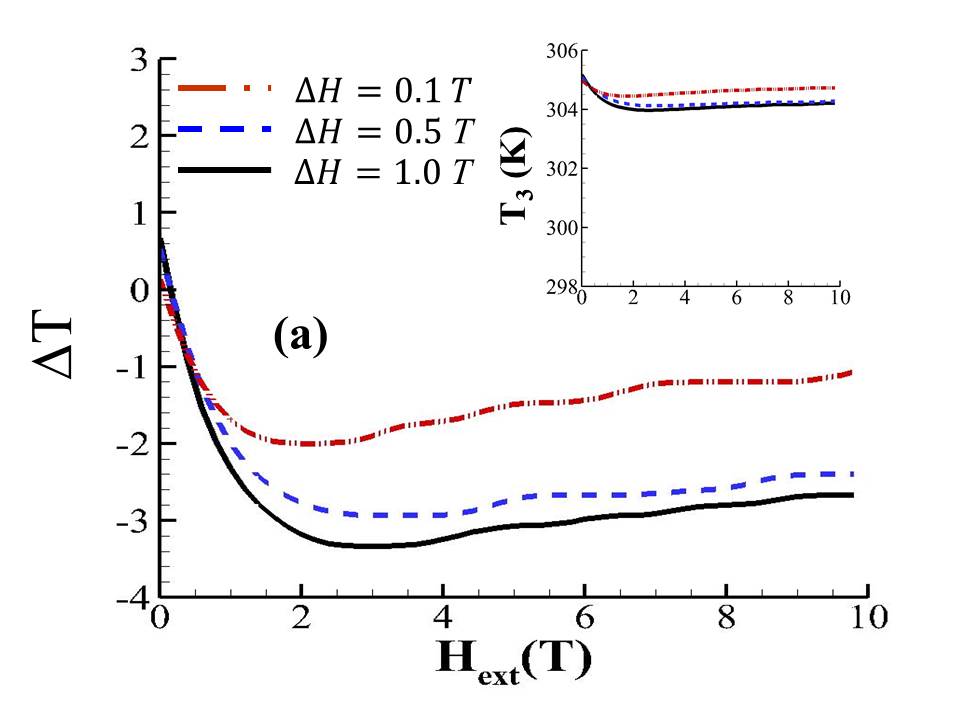}
\includegraphics[angle=0,scale=0.32,angle=0]{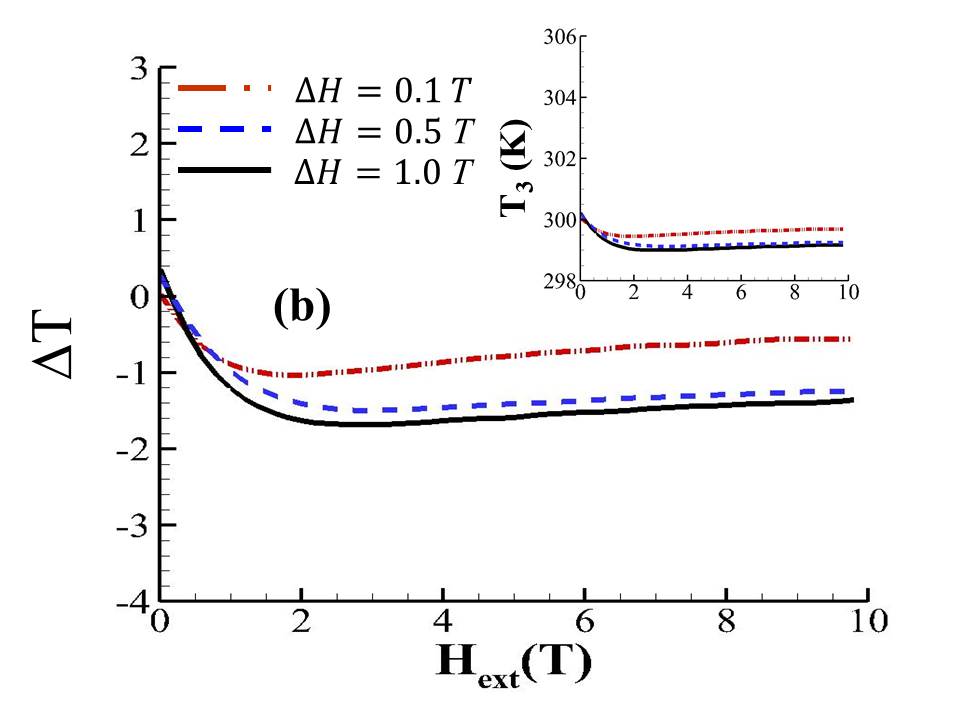}
\caption{(a) Latteral temperature gradient $\Delta T=T_3-T_4$ generated by ISTHE in a regular square lattice of InSb spherical nanoparticles ($R=50\;nm$) with respect to the magnitude of external magnetic field (applied in the $\bold{z}$-direction) for different gradient $\Delta H$ between particles $1$ and $2$ when the bath temperature is $T_b=300\;K$ and $T_1=T_2=T_b+10$. Inset: Temperature of particle $3$. (b) Latteral temperature gradient when $T_1=T_2=T_b$. Inset: Temperature of particle $3$. The side of square is $2\sqrt2 R$ (center to center).}
\end{figure}

Here below we show that this flux is directly related to the gradient of spin angular momentum (SAM) through the network and to its non-monotic evolution with respect to $H_{ext}$ making this effect, a thermal analog of the inverse spin Hall effect. To start, let us remind that the spectrum of the SAM density in the system is defined as~\cite{Bliokh}
\begin{equation}
\mathbf{S}_\omega(\bold{r})=\mathbf{S}^E_\omega(\bold{r})+\mathbf{S}^H_\omega(\bold{r}),\label{spin}
\end{equation}
where $\mathbf{S}^E_\omega(\bold{r})=\frac{\varepsilon_0}{2\omega}Im\langle \mathbf{E}^*(\bold{r})\times\mathbf{E}(\bold{r}) \rangle$ and $\mathbf{S}^H_\omega(\bold{r})=\frac{\mu_0}{2\omega}Im\langle \mathbf{H}^*(\bold{r})\times\mathbf{H}(\bold{r}) \rangle$, denote respectively the electric and magnetic contribution to the SAM. The local  electric and magnetic fields $ \mathbf{E}$ and $\mathbf{H}$ are related to fluctuating dipolar moments $\mathbf{p}_{j}^{fluc}$ as follows~\cite{pba2011,biehs1}
\begin{equation}
  \mathbf{E}(\bold{r})=\omega^{2}\mu_{0}\underset{j=1}{\overset{N}{\sum}}\mathds{G}^{EE}(\bold{r},\bold{r_j})\mathbf{p}_{j}^{fluc},
  \label{Eq:field_fluc}
\end{equation}
and
\begin{equation}
  \mathbf{H}(\bold{r})=i\omega\underset{j=1}{\overset{N}{\sum}}\mathds{G}^{HE}(\bold{r},\bold{r_j})\mathbf{p}_{j}^{fluc}.
  \label{Eq:mag_fluc}
\end{equation}
\begin{figure}
\centering
\includegraphics[angle=0,scale=0.31,angle=0]{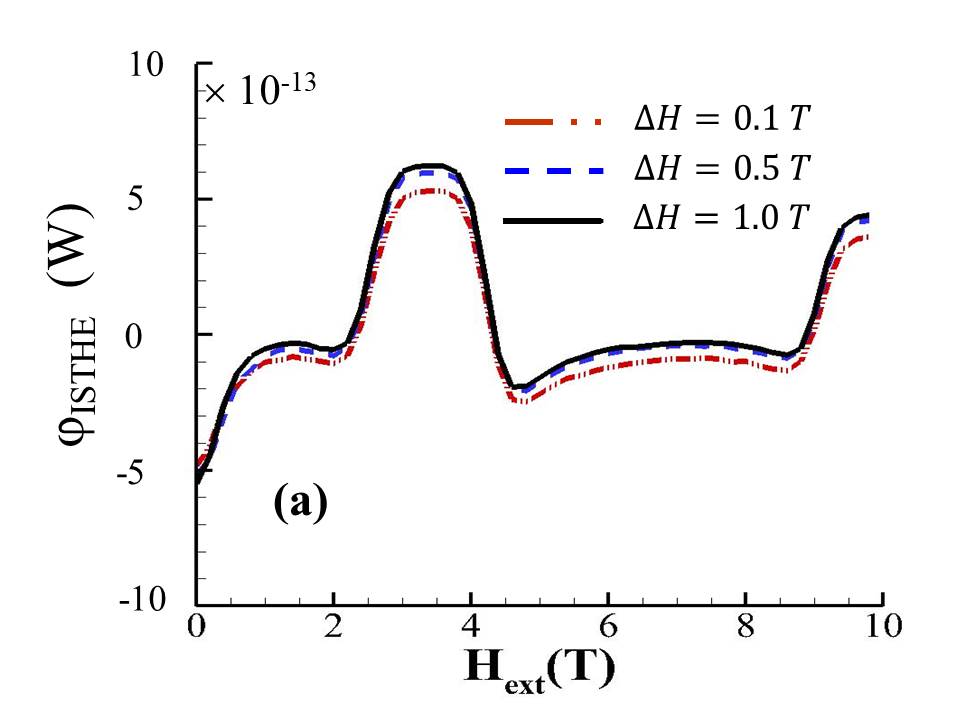}
\includegraphics[angle=0,scale=0.31,angle=0]{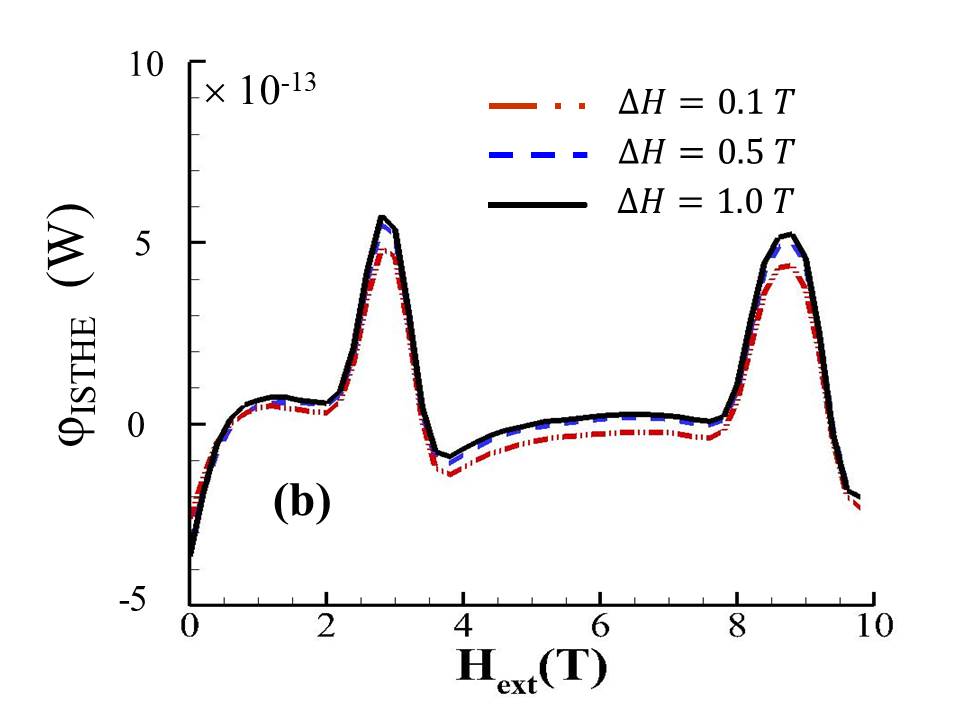}
\caption{Heat power exchanged by inverse thermal Hall effect between the particles $3$ and $4$  due to the  inverse spin thermal Hall effect in the same conditions as  in  (a) Fig.2(a) when $\delta T=10\: K$ and (b) Fig.2(b) where $\delta T=0$.}
\end{figure}
Here $ \mathds{G}^{HE}$ denotes the full magnetic-electric Green tensor defined as
\begin{equation}
  \mathds{G}^{HE}(\bold{r},\bold{r_j})=\underset{i=1}{\overset{N}{\sum}}\mathscr{G}_0^{HE}(\bold{r},\bold{r_i})\mathds{T} _{HE,ij}^{-1},
  \label{full_Green_magnetic}
\end{equation}
where $\mathscr{G}_0^{HE}$  is the magnetic-electric propagator in vaccum given by
\begin{equation}
 \mathscr{G}_0^{HE}(\bold{r'},\bold{r''})= \nabla\times\mathscr{G}_0^{EE}(\bold{r'},\bold{r''})\label{scat}
\end{equation}
while $\mathds{T} _{HE}$ is obtained by substituting $\mathscr{G}_0^{EE}\rightarrow\mathscr{G}_0^{HE}$ in expression~(\ref{Eq:A0}).
From relations (\ref{Eq:field_fluc}) and (\ref{Eq:mag_fluc}) and using the fluctuation dissipation theorem~\cite{Callen}
\begin{equation}
\langle p^{f}_{i,l}p_{j,n}^{f*}\rangle=\frac{\epsilon_{0}}{i\omega}\hat{\alpha}_{i,ln}\Theta_{\omega}(T_{i})\delta_{ij}\delta_{ln},
\label{FDT}
\end{equation}
\begin{figure}
\centering
\includegraphics[angle=0,scale=0.3,angle=0]{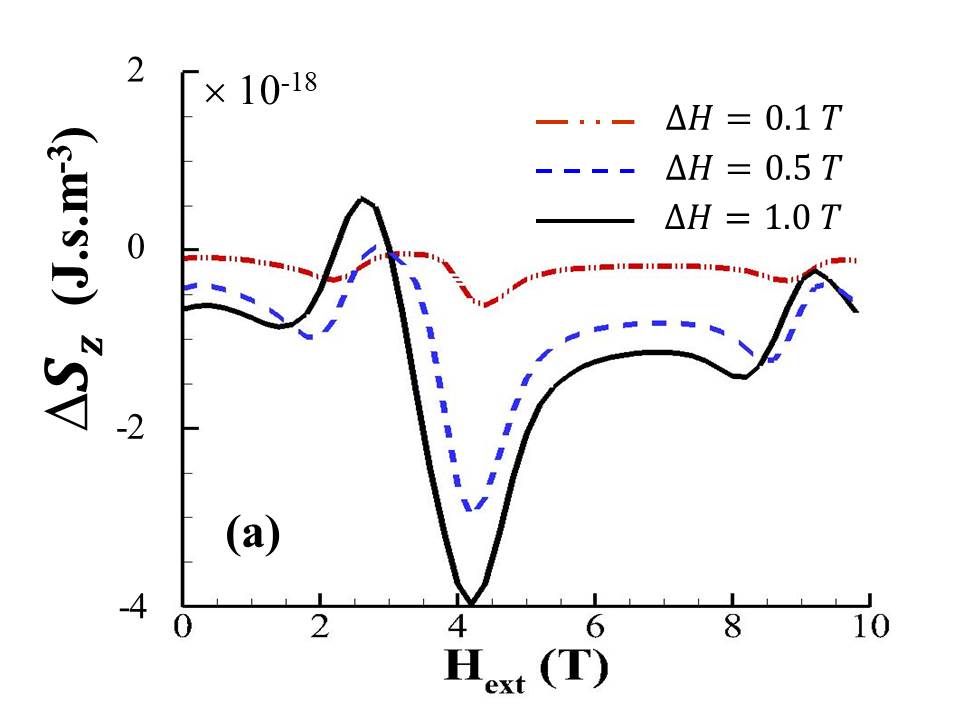}
\includegraphics[angle=0,scale=0.35,angle=0]{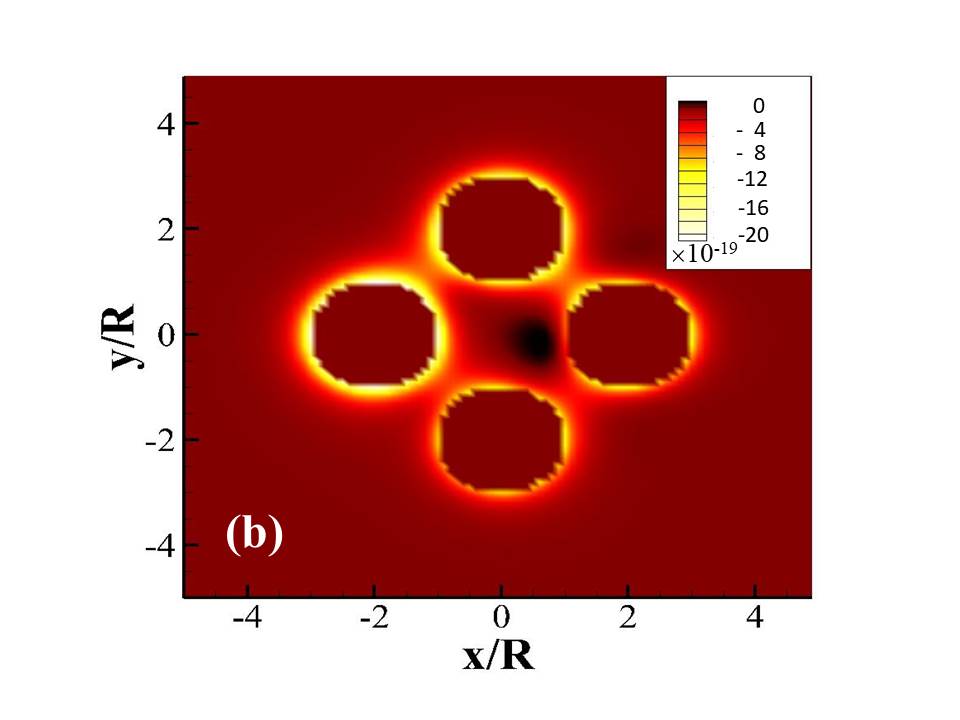}
\includegraphics[angle=0,scale=0.33,angle=0]{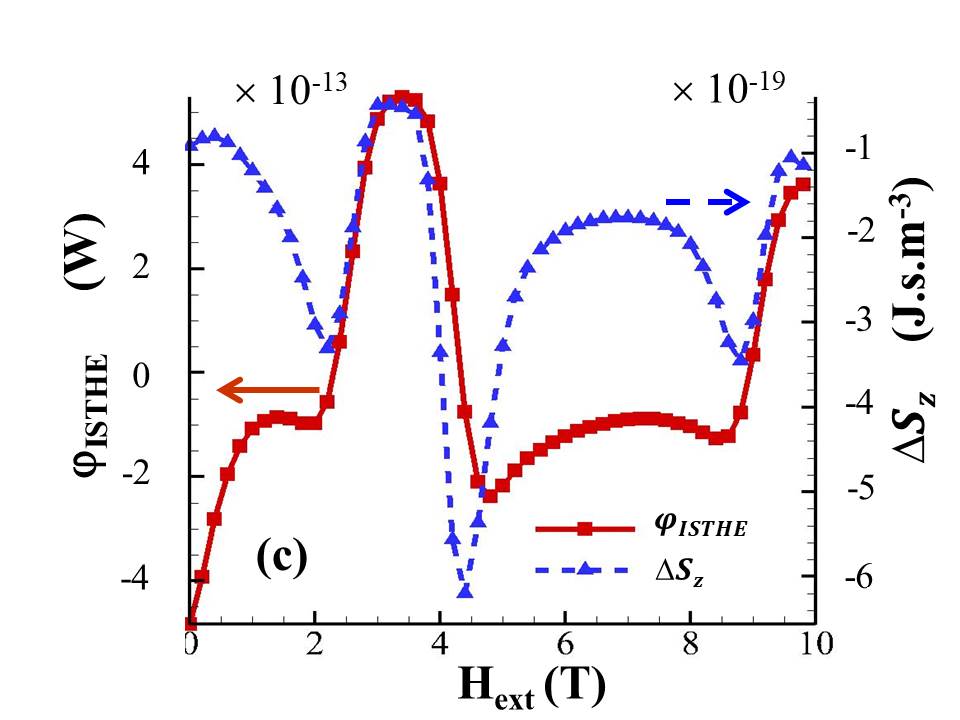}
\caption{(a) Variation $\Delta S_z=S_z^+-S_z^-$ of SAM between particle $2$ ($S_z^+$) and $1$ ($S_z^-$) at $x=3R$ and  $x=-3R$ with respect of external magnetic field for different gradient $\Delta H$ ($z=0$) when $\delta T=10$ and $T_b=300\: K$. (b) Spatial distribution of  the normal component $S_z$  of the  light spin angular momentum density in the plane $z=0$ around the same network of InSb nanoparticles  when $\delta T=10$, $H_{ext}=2.6\:T$ and $\Delta H=1\:T$.  (c) Comparison of $\varphi_{ISTHE}$ and $\Delta S_z$ when $\Delta H=0.1 T$.}
\end{figure}
it is straighforward to show that the electric and magnetic components of the SAM density read
\begin{equation}
\begin{split}
{S}^E_{i,\omega}(\bold{r})=\frac{\omega^2}{c^4}\eta_{ilk}\:\:\:\:\:\:\:\:\:\:\:\:\:\:\:\:\:\:\:\:\:\:\:\:\:\:\:\:\:\:\:\:\:\:\:\:\:\:\:\:\:\:\:\:\:\:\:\:\:\:\:\:\:\:\:\:\:\:\:\:\:\:\:\:\:\:\:\:\:\:\:\:\:\:\\
\times\sum_{j=1}^N \sum_{m=1}^3 Im[\mathds{G}^{EE*}_{lm}(\bold{r},\bold{r}_j)\mathds{G}^{EE}_{km}(\bold{r},\bold{r}_j)\hat{\alpha}_{j,mm}] \Theta_{\omega}(T_m),\label{spin_elec}
\end{split}
\end{equation}
\begin{equation}
\begin{split}
{S}^H_{i,\omega}(\bold{r})=-\frac{1}{c^2}\eta_{ilk}\:\:\:\:\:\:\:\:\:\:\:\:\:\:\:\:\:\:\:\:\:\:\:\:\:\:\:\:\:\:\:\:\:\:\:\:\:\:\:\:\:\:\:\:\:\:\:\:\:\:\:\:\:\:\:\:\:\:\:\:\:\:\\:\:\:\:\:\:\:\:\:\:\:\:\:\:\:\:\:\\
\times\sum_{j=1}^N \sum_{m=1}^3 Im[\mathds{G}^{HE*}_{lm}(\bold{r},\bold{r}_j)\mathds{G}^{HE}_{km}(\bold{r},\bold{r}_j)\hat{\alpha}_{j,mm}] \Theta_{\omega}(T_m),\label{spin_mag}
\end{split}
\end{equation}
where $\mathds{\eta}$ denotes the Levi-Cvita tensor and $\hat{\boldsymbol{\alpha}}_{j}\equiv\frac{(\boldsymbol{\alpha}_j-\boldsymbol{\alpha}^*_j)} {2i}$ is the antisymmetric part of polarizability tensor $\boldsymbol{\alpha}_j$.
In Fig.4(a) we see that the gradient of $\bold{H}_{ext}$  induces a non-trivial variation of SAM (i.e. $\mathbf{S}(\bold{r})=\int_{0}^{\infty}\frac{\rd\omega}{2\pi}\mathbf{S}_\omega(\bold{r})$) through the system. This "driving force" is directly responsible for the asymmetry in the temperature distribution inside the system as shown in Figs.2. The spatial distribution of the non-vanishing component $S_z$ of SAM plotted in Fig. 4(b), in the specific case where $H_{ext}=1\:T$ and $\Delta H=0.5\:T$,  shows the tiny asymmetry in the system between the upper and lower half planes. 
The variation in the heat flux exchanged between particles $3$ and $4$, as well as the change in the sign of the lateral temperature gradient with respect to the external magnetic field strength, arise from the involvement of different poles  ($\varepsilon_{1}+2=\pm \varepsilon_{2}$) of the nanoparticles during their coupling. As highlighted in~\cite{Ott1}, for an isolated magneto-optical particle, the angular momentum associated with these poles can be oriented either upward (counterclockwise) or downward (clockwise). Consequently, the strength and spatial distribution of the applied magnetic field determine which resonant modes predominantly contribute to the coupling. This explains why, at certain values of 
${H}_{ext}$, the temperature of particle $3$ can exceed or fall below that of particle $4$. It also accounts for the sign reversal observed in the heat power plotted in Fig. 3

To finish we demonstrate the connection between the variation in spin angular momentum and the flux induced by the inverse spin thermal Hall effect within the system. While the relationship between these two quantities is not easily expressed through their respective mathematical formulations, it can be revealed through a statistical analysis of their correlations. As shown in Fig. 4(c), the flux and the SAM gradient are not linearly related. However, both quantities exhibit a similar trend as a function of the external magnetic field $H_{ext}$. Specifically, as one increases, the other consistently increases, suggesting a nonlinear relationship between them with respect to $H_{ext}$.  To further explore this connection, we calculate the Spearman rank correlation coefficient~\cite{Gottfried}
\begin{equation}
\rho=1-\frac{6\underset{i=1}{\overset{N}{\sum}d^2_i}}{n(n^2-1)},\label{Spearman}
\end{equation}
where $d_i=R[\Delta S_z(H_i)]-R[\varphi_{ISTHE}(H_i)]$ is the difference between the two ranks $R$ of each set of data. For $\Delta H=0.1$ we obtain a correlation coefficient of  $\rho\approx 0.7$ over the range $H_{ext}\in[2;10]$ indicating a strong positive correlation between the inverse spin thermal Hall flux and the variation of SAM. However, this correlation coefficient decreases to $\rho=0.35$ when data from lower magnetic fields are included. This decay is related to the significant reduction in SAM at low $H_{ext}$ i
values, where the particles tend to become isotropic. Additionally, the lack of a perfect correlation (i.e. $\rho\sim 1$) and its degradation can be explained by the fact that the SAM variation along the $\bold{x}$ axis is measured between two arbitrary points (at $x=\pm 3R$ in Fig.4(c)) on opposite sides of particles $1$ and $2$, rather than considering the full distribution of SAM values along this axis. 

In summary, an ISTHE effect has been predicted in non-reciprocal networks  due to a symmetry breaking induced by the gradient of spin angular momentum of thermal photons. This effect
could find direct applications in the field of thermal management. The development of spin-based heat engines is another promising avenue exploiting spin angular momentum variations driven by magnetic field gradients. Finally the dynamic control of spin could find applications for pyroelectric energy conversion.

\begin{acknowledgements}
This work was supported by the French Agence Nationale de la Recherche (ANR), under grant ANR-21-
CE30-0030 (NBODHEAT). The author acknowledges discussions with S.-A. Biehs.
\end{acknowledgements}

\end{document}